\documentstyle[12pt,sidefig,psfig,epsf]{article}

\catcode`\@=11
\long\def\@makefntext#1{
\protect\noindent \hbox to 3.2pt {\hskip-.9pt

$^{{\ninerm\@thefnmark}}$\hfil}#1\hfill}                %CAN BE USED

\def\@makefnmark{\hbox to 0pt{$^{\@thefnmark}$\hss}}  %ORIGINAL

\def\ps@myheadings{\let\@mkboth\@gobbletwo
\def\@oddhead{\hbox{}
\rightmark\hfil\ninerm\thepage}
\def\@oddfoot{}\def\@evenhead{\ninerm\thepage\hfil
\leftmark\hbox{}}\def\@evenfoot{}
\def\sectionmark##1{}\def\subsectionmark##1{}}

\renewcommand{\thefootnote}{\fnsymbol{footnote}}

\newcounter{sectionc}\newcounter{subsectionc}\newcounter{subsubsectionc}

\renewcommand{\section}[1] {\vspace*{0.6cm}\addtocounter{sectionc}{1}
\setcounter{subsectionc}{0}\setcounter{subsubsectionc}{0}\noindent
        {\normalsize\bf\thesectionc. #1}\par\vspace*{0.4cm}}

\renewcommand{\subsection}[1] {\vspace*{0.6cm}\addtocounter{subsectionc}{1}
        \setcounter{subsubsectionc}{0}\noindent
        {\normalsize\it\thesectionc.\thesubsectionc. #1}\par\vspace*{0.4cm}}

\renewcommand{\subsubsection}[1]
    {\vspace*{0.6cm}\addtocounter{subsubsectionc}{1}
     \noindent {\normalsize\rm\thesectionc.\thesubsectionc.\thesubsubsectionc.
        #1}\par\vspace*{0.4cm}}

\newcounter{appendixc}

\newcounter{subappendixc}[appendixc]

\newcounter{subsubappendixc}[subappendixc]

\renewcommand{\appendix}[1] {\vspace*{0.6cm}
        \refstepcounter{appendixc}
        \setcounter{figure}{0}
        \setcounter{table}{0}
        \setcounter{equation}{0}
        \renewcommand{\thefigure}{\Alph{appendixc}.\arabic{figure}}
        \renewcommand{\thetable}{\Alph{appendixc}.\arabic{table}}
        \renewcommand{\theappendixc}{\Alph{appendixc}}
        \renewcommand{\theequation}{\Alph{appendixc}.\arabic{equation}}
        \noindent{\bf Appendix \theappendixc #1}\par\vspace*{0.4cm}}

\def\abstracts#1{{
        \centering{\begin{minipage}{12.2truecm}\footnotesize\baselineskip=12pt\noindent
        \parindent=0pt #1
        \end{minipage}}\par}}

\renewenvironment{thebibliography}[1]
       {\begin{list}{\arabic{enumi}.}
        {\usecounter{enumi}\setlength{\parsep}{0pt}
\setlength{\leftmargin 1.25cm}{\rightmargin 0pt}
         \setlength{\itemsep}{0pt} \settowidth
        {\labelwidth}{#1.}\sloppy}}{\end{list}}

\newcounter{itemlistc}
\newcounter{romanlistc}
\newcounter{alphlistc}
\newcounter{arabiclistc}

\newcommand{\fcaption}[1]{
        \refstepcounter{figure}
        \setbox\@tempboxa = \hbox{\footnotesize Fig.~\thefigure. #1}
        \ifdim \wd\@tempboxa > 6in
           {\begin{center}
        \parbox{6in}{\footnotesize\baselineskip=12pt Fig.~\thefigure. #1}
            \end{center}}
        \else
             {\begin{center}
             {\footnotesize Fig.~\thefigure. #1}
              \end{center}}
        \fi}

\newcommand{\tcaption}[1]{
        \refstepcounter{table}
        \setbox\@tempboxa = \hbox{\footnotesize Table~\thetable. #1}
        \ifdim \wd\@tempboxa > 6in
           {\begin{center}
        \parbox{6in}{\footnotesize\baselineskip=12pt Table~\thetable. #1}
            \end{center}}
        \else
             {\begin{center}
             {\footnotesize Table~\thetable. #1}
              \end{center}}
        \fi}

\def\@citex[#1]#2{\if@filesw\immediate\write\@auxout
        {\string\citation{#2}}\fi
\def\@citea{}\@cite{\@for\@citeb:=#2\do
        {\@citea\def\@citea{,}\@ifundefined
        {b@\@citeb}{{\bf ?}\@warning
        {Citation `\@citeb' on page \thepage \space undefined}}
        {\csname b@\@citeb\endcsname}}}{#1}}

\newif\if@cghi
\def\cite{\@cghitrue\@ifnextchar [{\@tempswatrue
        \@citex}{\@tempswafalse\@citex[]}}
\def\citelow{\@cghifalse\@ifnextchar [{\@tempswatrue
        \@citex}{\@tempswafalse\@citex[]}}
\def\@cite#1#2{{$\null^{#1}$\if@tempswa\typeout
        {IJCGA warning: optional citation argument
        ignored: `#2'} \fi}}

 1
 1
 1

\font\ninerm=cmr9

\begin{document}

\centerline{\large\bf Zn-related deep centers in wurtzite GaN}

\baselineskip=22pt

\baselineskip=16pt
%\vfill
\vspace*{0.2cm}
\centerline{\normalsize Fabio Bernardini,$^1$ Vincenzo Fiorentini,$^1$ 
and R.~M.~Nieminen$^2$}
\vspace*{0.1cm}
\baselineskip=13pt
\begin{center}
%\centerline
{\footnotesize\it (1) Istituto Nazionale di Fisica della Materia --
 Dip. Scienze Fisiche,
Universit\`a di Cagliari, Italy}\\
\baselineskip=12pt
%\centerline{\footnotesize E-mail: fabio@alfis.unica.it, fiore@alfis.unica.it}
%\vspace*{0.3cm}
%\centerline{\footnotesize and}
%\vspace*{0.3cm}
%\centerline{\footnotesize   }
\baselineskip=13pt
%\centerline
{\footnotesize\it (2) 
Laboratory of Physics, Helsinki University
 of Technology, FIN-02150 Espoo, Finland} 
%\centerline{\footnotesize E-mail: rni@hugo.hut.fi}
\end{center}

%\vfill
\vspace*{0.5cm}

\abstracts{Zn in GaN forms an efficient radiative center and acts as a
deep acceptor which can make the crystal insulating. Four different 
Zn-related centers have been by now identified, leading to light
emission in the range between 1.8 eV and 2.9 eV. 
We present a first-principles investigation 
total energy and electronic structure calculations for Ga-substitutional
and hetero-antisite N-substitutional Zn in wurtzite GaN, using ultrasoft
pseu\-do\-po\-ten\-tials and a conjugate-gradient total energy minimization 
method. Our results permit the identification of the blue-light
emission center as the substitutional acceptor, while contrary to a
common belief the Zn$_{\rm N}$ heteroantisite  has a very high formation
energy and donor behavior, which seems to exclude it as the
origin of the other centers.
}

%\vspace*{0.6cm}
\normalsize\baselineskip=15pt
\setcounter{footnote}{0}
\renewcommand{\thefootnote}{\alph{footnote}}

\section{Introduction}
Wide-band-gap nitride semiconductors are promising 
material for the fabrication of blue-light emitting 
diode lasers.
As-grown GaN is commonly found to be {\it n}-type conductive due to
 native donors;% commonly associated with the nitrogen vacancy.
\,\cite{Perlin.PRL,Maruska.APL,Ilegems.JPCS}
Zn doping has been succesfully 
employed to obtain semi-insulating material.~\cite{Nakamura.APL}
There is general agreement that Zn is  a very efficient 
donor compensator,
 that it introduces at least one deep acceptor level,
and that it can act as recombination center.
In recent works,\,\cite{Teisseyre,Bergman.JAP} for moderate 
doping concentrations,
only a single emission 
peaking at $\simeq$ 2.9 eV
has been detected in Zn-doped GaN. Other authors \cite{Monemar.JAP,Jacob.JCG} 
report four luminescence bands peaking at 
 2.9, 2.6, 2.2 and 1.8 eV (referred to as bands A, B, C and D)
 at higher Zn concentrations. 
While the blue-light emission (band A)
seems clearly related to a deep acceptor level of  substitutional Zn,
there is some debate about the origin of the other emission centers. 
Among the possible candidates, 
the most  widely credited   by the experimentalists
are the acceptor levels originating from the various charge states 
with heteroantisite impurities Zn$_{\rm N}$.
 %are Zn-nitrogen vacancies complexes and .
Monemar  {\it et al.}~\cite{Monemar.JAP}
suggest for the identity of levels B-D the charge states 
of Zn$_{\rm N}$, namely Zn$^-_{\rm N}$,
Zn$^{2-}_{\rm N}$ and Zn$^{3-}_{\rm N}$. 
Within this model,  Zn$_{\rm N}$ should be able to bind up to three
electrons with  associated binding energies of  0.65, 1.02 and 1.43 eV.
\cite{Monemar.JAP}

Recent theoretical investigations on  native defects
 in GaN\,\cite{Neugebauer.RC} have  shown
that the  Ga$_{\rm N}$ antisite is 
a donor for a wide electron chemical potential range,
and that its   formation energy  is so high as to allow
only negligible concentrations  in 
equilibrium conditions.
In view of the similar electronic structure and atomic size
of  Ga and Zn, the behavior of the Zn heteroantisite should not be 
too different from that of the antisite proper; this points against the 
commonly accepted identification of levels B-D with Zn$_{\rm N}$ 
acceptor levels. Of course,  a direct first-principles 
investigation may give important clues
to   confirm or refute
 this argument. Such calculation is provided in the present paper.
Our calculations confirm 
that Zn$_{\rm Ga}$ is a single acceptor with a
thermal ionization energy of 0.33 eV, and a recombination center with
emission at 2.95 eV; further, they rule out the
possibility that the B, C, and D  emission bands may be related to 
Zn$_{\rm N}$, because of the high formation energy for this defect and
 of the gap level positions.

\section{Method}

Total energies and forces are calculated within local density
functional theory, using a conjugate-gradient 
minimization scheme,\cite{ksv} plane-waves,
and  ultrasoft 
pseu\-do\-potentials.\cite{vand} 
These  allow for an  accurate description of the 
localized Ga and Zn 3$d$, and N 2$p$ electrons using 
 a cutoff of only 25 Ryd. Full defect geometry optimization 
is performed using a newly developed method (details given elsewhere),
in  32-atom wurtzite GaN supercells.
The formation energy of the defect in a charge state {\it q} is given by 
\begin{equation}
E_f(q) = 
E^{tot}(q)  -n^{\rm Ga} \mu^{\rm Ga}   -n^{\rm N} \mu^{\rm N}-n^{\rm Zn} 
\mu^{\rm Zn} + q E_{\rm F},
\label{eform}
\end{equation}
where
$E^{tot}(q)$ is the defected supercell total energy in a specific
 charge state, $E_{\rm F}$ the electron chemical potential (the Fermi 
level), $n^{\rm Ga}$,  $n^{\rm N}$, and $n^{\rm Zn}$
are the number of Ga,  N, and Zn atoms, $\mu^{\rm Ga}$, $\mu^{\rm N}$, and 
$\mu^{\rm Zn}$ their chemical potentials.
These chemical potentials are not independent as they must satisfy the 
equilibrium conditions with GaN and Zn$_3$N$_2$ compounds:
\begin{equation}
\mu^{\rm GaN}      = \mu^{\rm Ga} + \mu^{\rm N},
~~~~~\mu^{\rm Zn_3 N_2} = 3\mu^{\rm Zn} + 2\mu^{\rm N}. 
\end{equation}
Using these relations and Eq.~\ref{eform} we get 
\begin{equation}
E_f(q)= E^{tot}(q) - n^{\rm Ga}\mu^{\rm GaN}-\frac{1}{3}\mu^{\rm Zn_3
N_2} - (n^{\rm N} - n^{\rm Ga} - 2/3)~\mu^{\rm N} + q E_{\rm F}  
\end{equation}
where $\mu^{\rm N}$ is allowed to vary in the interval
\begin{equation}
\mu^{\rm N_2} \geq \mu^{\rm N}  \geq  \mu^{\rm N_2} 
+ \frac{1}{2} \Delta H^{\rm Zn_3 N_2}  
\end{equation}
Thus the degrees of freedom for the formation energy are 
reduced to one, the nitrogen chemical potential.
It should be noted that in N-poor conditions the extremal 
value for $\mu^{\rm N}$ is determined
by the Zn$_3$N$_2$ formation 
\begin{table}[t]
\tcaption{Zn impurities in wurtzite GaN.  
Formation energies E$_{\rm f}$ in N-rich and N-poor conditions,
relaxation contribution to  E$_{\rm f}$, change in 
distance to neighbors in the {\it a}-plane ($\Delta$d$_{a}$) and 
along the {\it c}-axis 
($\Delta$d$_{c}$) in percentage of the ideal bond lenght, are shown.
All energies are in  eV.
}
\label{tab:exp}
\small
\begin{center}
\begin{tabular}{|l|r|r|r|r|r|}\hline\hline
Defect      & E$_{\rm f}$ N-rich & E$_{\rm f}$ N-poor & E$_{\rm r}$ & 
$\Delta$ d$_{\rm a}$ & $\Delta$ d$_{\rm c}$\\ \hline
Zn$^{-}_{\rm Ga}$ &  1.54     &   1.75     &  0.14      &  +1.1 & +2.2  \\ 
Zn$^0_{\rm Ga}$   &  1.21     &   1.42     &  0.03      &  +1.1 & --0.5  \\ 
\hline
Zn$^{2-}_{\rm N}$ & 13.15     &   12.09    &  2.64      & +10.2 & +9.3  \\ 
Zn$^{1-}_{\rm N}$ &  9.68     &    8.62    &  3.26      & +13.0 & +9.9  \\ 
Zn$^{0}_{\rm N}$  &  7.66     &    6.60    &  3.22      & +11.7 & +12.1 \\ 
Zn$^{1+}_{\rm N}$ &  5.44     &    4.38    &  4.09      & +13.5 & +17.0 \\    
Zn$^{2+}_{\rm N}$ &  4.87     &    3.81    &  5.14      & +16.1 & +18.0 \\
\hline
Mg$^{0}_{\rm Ga}$ &  1.40     &    1.95    &  0.21      & +3.2  & +3.2  \\
Ca$^{0}_{\rm Ga}$ &  2.15     &    2.70    &  1.87      & +10.0 & +12.7 \\
\hline\hline
\end{tabular}
\end{center}
\end{table}
energy (1.28 eV) instead of that of GaN. This affects appreciably
the formation energies 
of the substitutional and heteroantisite defect:
 the range of variation for the formation energy will be
only 0.21 eV for Zn$_{\rm Ga}$, a
much narrower range than in the case of native defects, 
 and 1.06 for Zn$_{\rm N}$. All the chemical potentials are calculated ab 
initio with compatible technical ingredients.

\section{Results and Discussion}
In Table 1, we summarize the results for Zn,
compared  with data for the Mg and Ca acceptors (discussed elsewhere in 
these Proceedings). In Fig. 1 we display the defect formation energies 
vs. the Fermi energy in the N-rich case. 
The results for Zn$_{\rm Ga}$ confirm the identification
of this defect with the blue-light recombination center.
 Zn$_{\rm Ga}$ is a single acceptor with a thermal
ionization energy of 0.33 eV, in good agreement with 
the existing experimental data.
The calculated formation energy for Zn$^0_{\rm Ga}$ 
is quite low,
which is compatible with the high Zn doping 
concentrations achievable in GaN.
Such a  low formation cost matches the
size similarities  of the Zn and Ga atoms.
Zn is able to occupy the substitutional site in the GaN lattice
without inducing a significant distortion;  Mg, and especially 
Ca, have a bigger
atomic radius and their incorporation causes sizable relaxations.
Increasing atomic size mismatch correlates (see Table 1) with
increasing relaxation energy and formation energy.
\begin{sidefigure}
\unitlength=1cm
\vspace{-2.5cm}\hspace{-0.6cm}
\begin{picture}(7.5,7.0)
\put(0.0,0.2){\psfig{figure=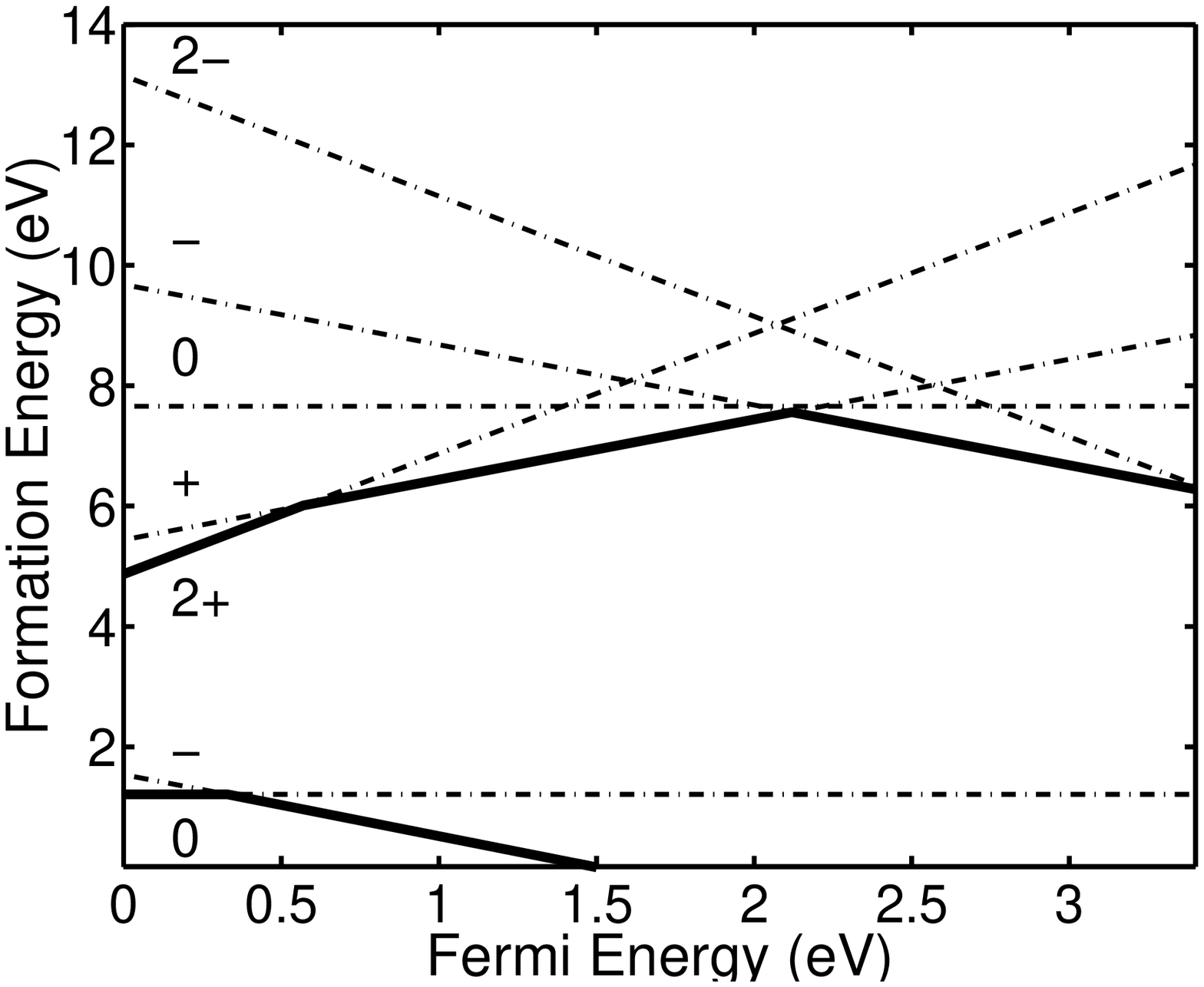,width=6.6cm,height=4.6cm}}
\end{picture}
\parbox{6.5cm}{\footnotesize\baselineskip=12pt Fig.~1:
~Defect formation energy vs. Fermi energy E$_F$,
 referred to  the valence band top.}
\end{sidefigure}
We found the Frank-Condon shift for Zn$^-_{\rm Ga}$/Zn$^0_{\rm Ga}$
to be 0.12 eV (very close to the difference of the
relaxation energies of  the two states). Adding this to the
thermal binding energy gives  0.45 eV for the optical ionization level.
This is in remarkable agreement with  the experimental  luminescence
 emission energy for band A of  2.9 eV which (using 
the experimental value of 3.4 eV for the   GaN gap) yields an optical 
level at 0.5 eV  above the valence band top.
In Fig.~2 we plot the electronic density 
of the highest occupied state of the Zn$^-_{\rm Ga}$ defect 
along the atomic chains in the ($\bar{1}2\bar{1}0$) plane.
It is clear that the level is mainly of
 N $p_{xy}-$like character, with Zn 3$d$ admixture,
while components from N $p_z$ and Ga are
negligible. This feature is peculiar of {\it wurtzite} GaN, whose 
highest valence band is a N $p_{xy}$ doublet laying in the {\it a}-plane.

\vspace{1cm} 
\begin{sidefigure}
\unitlength=1cm
\vspace{-2.5cm}\hspace{-0.6cm}
\begin{picture}(7.5,7.0)
\put(0.0,0.2){\psfig{figure=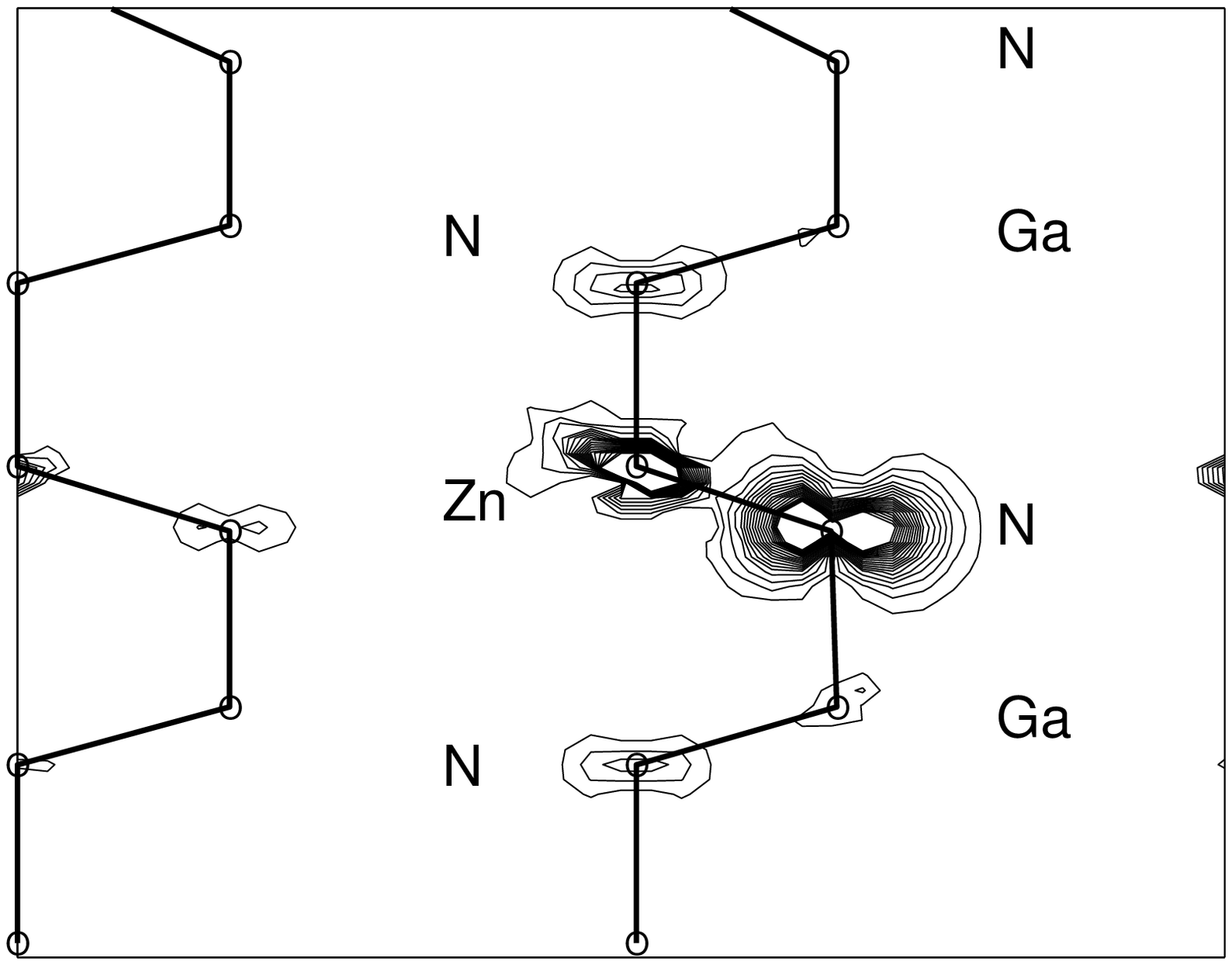,width=6.6cm,height=4.6cm}}
\end{picture}
\parbox{6.5cm}{\footnotesize\baselineskip=12pt Fig.~2:
Electronic density for the highest occupied state of Zn$^{-}_{\rm Ga}$
defect. }
\end{sidefigure}
%\hspace{-\parindent}
For  the Zn heteroantisite, the formation 
energy is very high. This rules out the possibility of
having appreciable concentration of this defect in equilibrium conditions. 
Further,  we find only two ionization
levels in the gap,
 $\epsilon(2+/+) = 0.57$ eV and
 $\epsilon(+/-) =  2.22$ eV. 
The doubly and singly positive, and singly negative
states are then the only ones  permitted; note the  ne\-ga\-ti\-ve-U 
behavior  causing the instability of the neutral cha\-rge state. This 
in contrast
with the  three levels (--,2--,3--) needed to explain the B, 
C, and D emission bands. The large formation energy  is related to 
the large ato\-mic size mismatch between N and Zn.
A strong outward relaxation occurs at the defect, of
a magnitude comparable with that reported for the Ga 
antisite.\,\cite{Neugebauer.RC}
We mention further that preliminary results on the 
formation energy of  interstitial Zn
indicate that this center should be present in very low concentrations
only, and cannot be responsible for the observed emission bands.
Further interesting preliminary results are available on Cd. The 
thermal ionization energy is 0.65, and the optical level is
0.79 eV.  
Cd therefore lends itself as a recombination center at
around  2.6 eV. This
may be of  use  in yellow-green emitters
based on InGaN alloys, where a lower In mole fraction may be needed,
with ensuing improved material quality.
Cd$_{\rm N}$ heteroantisites behave 
similarly to
Zn$_{\rm N}$, and are therefore irrelevant.

\end{document}